\documentclass[%
 reprint,
 amsmath,amssymb,
 aps,
 pre,
 nofootinbib
]{revtex4-2}

\usepackage{color}
\usepackage{amsfonts}
\usepackage{amssymb}
\usepackage{amsmath}
\usepackage{amsthm}
\usepackage{latexsym}
\usepackage{graphicx}
\usepackage{tabularx}
\usepackage{bm}
\usepackage{bbm}
\usepackage{array}
\usepackage{esvect}
\usepackage{import}
\usepackage{wrapfig}
\usepackage{csquotes}
\usepackage[makeroom]{cancel}
\usepackage[font=footnotesize,labelfont=bf]{caption}
\usepackage{appendix}

\newcommand\tab[1][6mm]{\hspace*{#1}}

\newcommand{\comment}[1]{}

\usepackage{scalerel}
\def\shrinkage{-2.4mu}
\def\vecsign#1{\rule[1.388\LMex]{\dimexpr#1-2.5pt}{.36\LMpt}%
	\kern-6.0\LMpt\mathchar"017E}
\def\dvecsign#1{\rule{0pt}{7\LMpt}\smash{\stackon[-1.989\LMpt]{%
			\SavedStyle\mkern-\shrinkage\vecsign{#1}}%
		{\rotatebox{180}{$\SavedStyle\mkern-\shrinkage\vecsign{#1}$}}}}
\def\dvec#1{\ThisStyle{\setbox0=\hbox{$\SavedStyle#1$}%
		\def\useanchorwidth{T}\stackon[-4.2\LMpt]{\SavedStyle#1}{\,\dvecsign{\wd0}}}}
\usepackage{stackengine,amsmath}
\stackMath

\begin{document}

\preprint{APS/123-QED}

\title{Self-propelled deformable particle model for keratocyte galvanotaxis}

\author{Ifunanya Nwogbaga$^1$}
\author{Brian A. Camley$^{1,2}$}%
\affiliation{%
 $^1$Department of Biophysics, Johns Hopkins University, Baltimore, MD 21218\\
 $^2$Department of Physics \& Astronomy, Johns Hopkins University, Baltimore, MD 21218
}%

\begin{abstract}
During wound healing, fish keratocyte cells undergo galvanotaxis where they follow a wound-induced electric field. In addition to their stereotypical persistent motion, keratocytes can develop circular motion without a field or oscillate around the field direction. We developed a coarse-grained phenomenological model that captures these keratocyte behaviors. We fit this model to experimental data on keratocyte response to an electric field being turned on. A critical element of our model is a tendency for cells to turn toward their long axis, arising from a coupling between cell shape and velocity, which gives rise to oscillatory and circular motion. Galvanotaxis is influenced not only by the field-dependent responses, but also cell speed and cell shape relaxation rate. When the cell reacts to an electric field being turned on, our model predicts that stiff, slow cells react slowly but follow the signal reliably. Cells that polarize and align to the field at a faster rate react more quickly and follow the signal more reliably. When cells are exposed to a field that switches direction rapidly, cells follow the average of field directions, while if the field is switched more slowly, cells follow a \enquote{staircase} pattern.

\end{abstract}

\maketitle

\section{\label{sec:intro}Introduction}

Directed cell migration is necessary for processes like wound healing, immune response, and developmental biology \cite{sengupta2021principles}. In healing wounds, epidermal keratocyte cells in fish and amphibians can follow electrical fields \cite{kunzenbacher1982dynamics,keren2008biophysical,radice1980locomotion}. It has been known since the 19$^\textrm{th}$ century that skin wounds generate electric fields \cite{du1849untersuchungen,mccaig2005controlling}. {\it In vivo,} injuries are believed to create changes in the ionic content, altering the transepithelial potential \cite{dietz1967roles,mycielska2004cellular} and subsequently leading to electric fields and migration \cite{kennard2020osmolarity}. {\it In vitro}, many single cells also have strong directional response to electric fields.  
Dictyostelium discoideum, human mesenchymal stem cells, fibroblasts, and keratocytes in electric fields typically migrate towards the cathode \cite{li2018electric,banks2015effects,sugimoto2012optimum,sun2013keratocyte,mycielska2004cellular,cooper1986motility}. Keratocytes in particular are relatively geometrically simple and have become a useful model system for cell motility in response to electric fields \cite{lee2020modeling,camley2017crawling,allen2020cell,mogilner2020experiment}. Interestingly, in addition to their stereotyped persistent motility, keratocytes also display spontaneous circular motion, a persistent turning  driven by asymmetries in traction forces and myosin contractility at the rear of the cell \cite{allen2020cell,lee2020modeling,mogilner2020experiment}. This behavior is also seen in detailed molecular models resolving cell shape and polarity \cite{camley2017crawling,allen2020cell,lee2020modeling,nickaeen2017free,camley2013periodic} and emerges from self-propelled deformable particle models \cite{ohta2009deformable}. In the presence of a field, keratocytes may oscillate around the field direction \cite{allen2013electrophoresis}.

The molecular details of galvanotaxis in keratocytes are not yet clear.  There is evidence that galvanotaxis is driven by electrophoresis of mobile, charged transmembrane proteins \cite{allen2013electrophoresis,mogilner2020experiment,kobylkevich2018reversing}, but the specific protein species acting as a sensor is unknown. In addition, the details of how this spatial asymmetry on the cell surface leads to initiation of the downstream signaling cues and migration are not fully understood. Despite this uncertainty, it is clear that electric fields are a key cue in wound healing, overriding other signals \cite{kennard2020osmolarity,zhao2009electrical}. As the study of directed motion in electric fields is expanded to the study of groups of cells \cite{lalli2015collective,sun2020pi3k,camley2018collective,li2012cadherin,zajdel2020scheepdog,shim2021overriding,dawson2021computational}, there is a need for phenomenological models that can quantitatively describe galvanotaxis.

We developed a stochastic model of single-cell galvanotaxis describing the cell as a self-propelled deformable particle \cite{hiraiwa2010dynamics,hiraiwa2013theoretical,ohta2009deformable,ohta2009deformationreaction,ohta2016simple}, tracking the cell's shape and velocity. Experimentally, cell shape plays a key role in motility. A cell's shape affects the traction forces it applies to the surface, subsequently affecting its motility \cite{zhong2013impact}. Keratocytes adopt different morphologies for migrating forward versus making turns  \cite{allen2020cell}. We therefore include a coupling between the cell's velocity and its shape to allow the shape to affect motility; this coupling is known to create circular motion \cite{ohta2009deformable}. In addition, we introduce a polarity angle to indicate the underlying biochemical polarization of the cell -- the ``compass" that responds to the electric fields. This polarity is then coupled with the cell's velocity. We show using linear stability analysis that, as in \cite{ohta2009deformable}, this model has regimes where persistent linear crawling by the cell is stable, and others where more complex behaviors like circular motion arise. We fit this model to the data of \cite{allen2013electrophoresis} studying how keratocyte directionality responds to an activated field. We then use these parameters to show how the cell's directionality depends on features like cell speed and the rate its shape relaxes to equilibrium. Finally, we show that the cell's rate of response to a switched-on electric field also can influence its response to rapidly-switched fields, where cells may reliably follow each field direction, follow the average of the fields, or develop a more complex behavior, depending on the cell's parameters.

\section{\label{sec:model}Model}
\begin{figure}[ht]
    \centering
	\includegraphics[width=8.65cm, height=4.95cm]{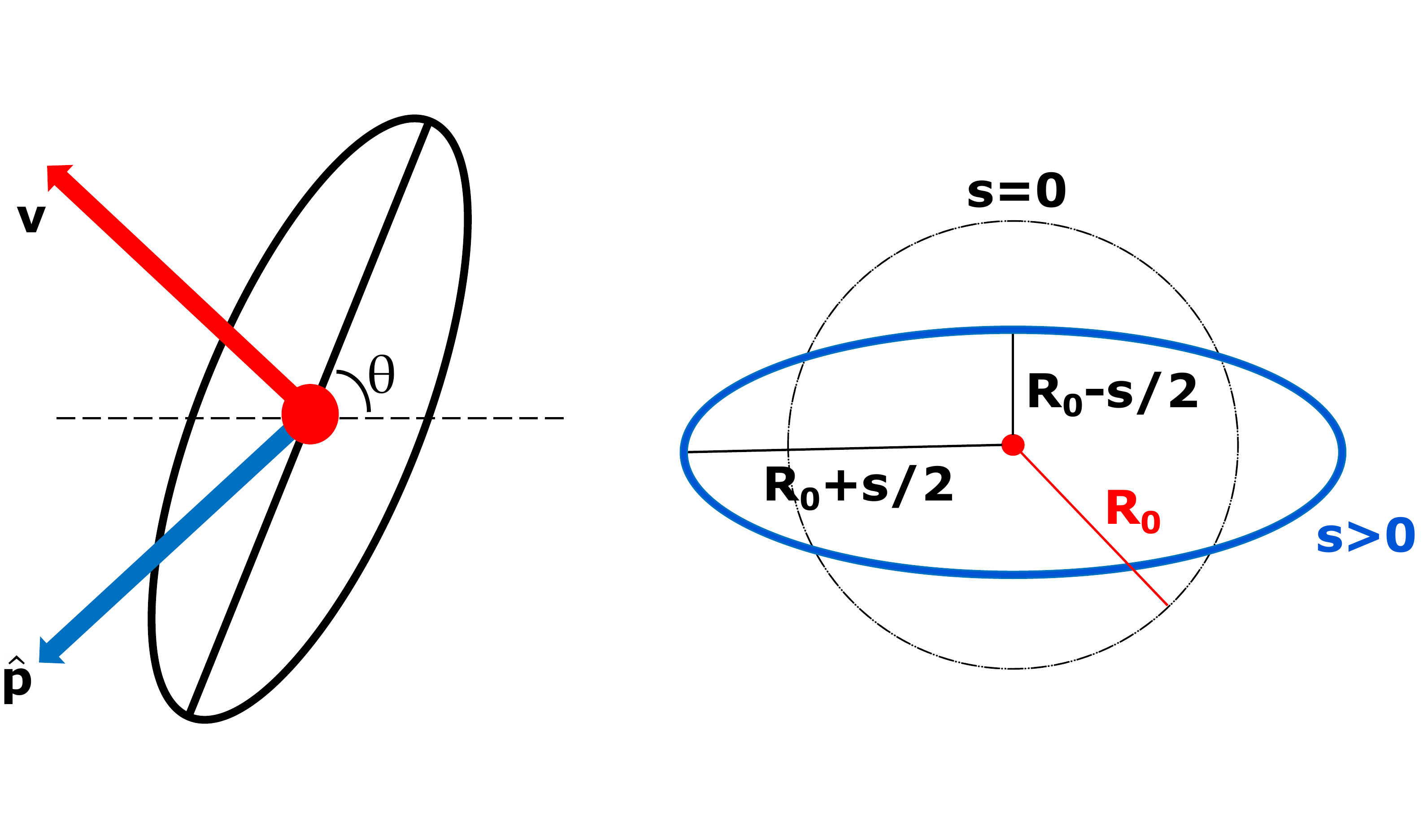}
	\caption{Cells are modeled as deformable ellipses (left), with a velocity $\mathbf{v}$, an underlying biochemical polarity unit vector $\hat{\mathbf{p}}$, Their orientation is characterized by a angle from the $\mathbf{\hat{x}}$-axis, $\theta$. Cells begin as circular (right) with radius $R_0$. They can deform to ellipses with semi-major and semi-minor axes lengths of $R_0\pm s/2$. s controls eccentricity, with $s=0$ denoting no deformation from a circle.}
	\label{fig:Ellipse}
\end{figure}

Our model includes four key variables to represent the state and motion of a keratocyte:  shape (characterized by a tensor $S_{\alpha\beta}$), center of mass $\mathbf{r}(t)$,  velocity $\mathbf{v} = \dfrac{\mathrm{d}\mathbf{r}}{\mathrm{d}t}$, and a internal polarity direction $\hat{\mathbf{p}}$ (Fig. \ref{fig:Ellipse}).  

The symmetric traceless shape tensor $S_{\alpha\beta}$ \cite{ohta2009deformable} captures the extent of deformation from a circle, $s(t)$, and the orientation of the cell's long axis, $\theta(t)$,
\begin{equation}
	\label{eq:tensor}
S_{\alpha\beta}=s\left(n_\alpha n_\beta-\dfrac{1}{2}\delta_{\alpha\beta}\right),
\end{equation}
where $\delta_{\alpha\beta}$ is the Kronecker delta and $\mathbf{\hat{n}}(t)=(\cos{\theta},\sin{\theta})$ points along the long axis of the cell (Fig. \ref{fig:Ellipse}). 
Here, $s(t)$ is a nonnegative number that describes the degree of deformation. $s=0$ is a circular cell while increasing $s$ increases cell eccentricity. The eigenvalues of $S_{\alpha\beta}$ are $\pm s/2$. We draw cells with a deformation $s$ as being an ellipse with the length of the semi-major and minor axes given by $A=R_0+s/2$ and $B=R_0-s/2$, where $R_0$ is the radius the cell would have if it were circular (Fig. \ref{fig:Ellipse}). We chose the deviation $s/2$ to be consistent with the definition of $s$ in the Fourier mode representation of \cite{ohta2009deformationreaction}.

\begin{figure}[ht]
    \centering
	\includegraphics[width=8.5cm, height=7.7cm]{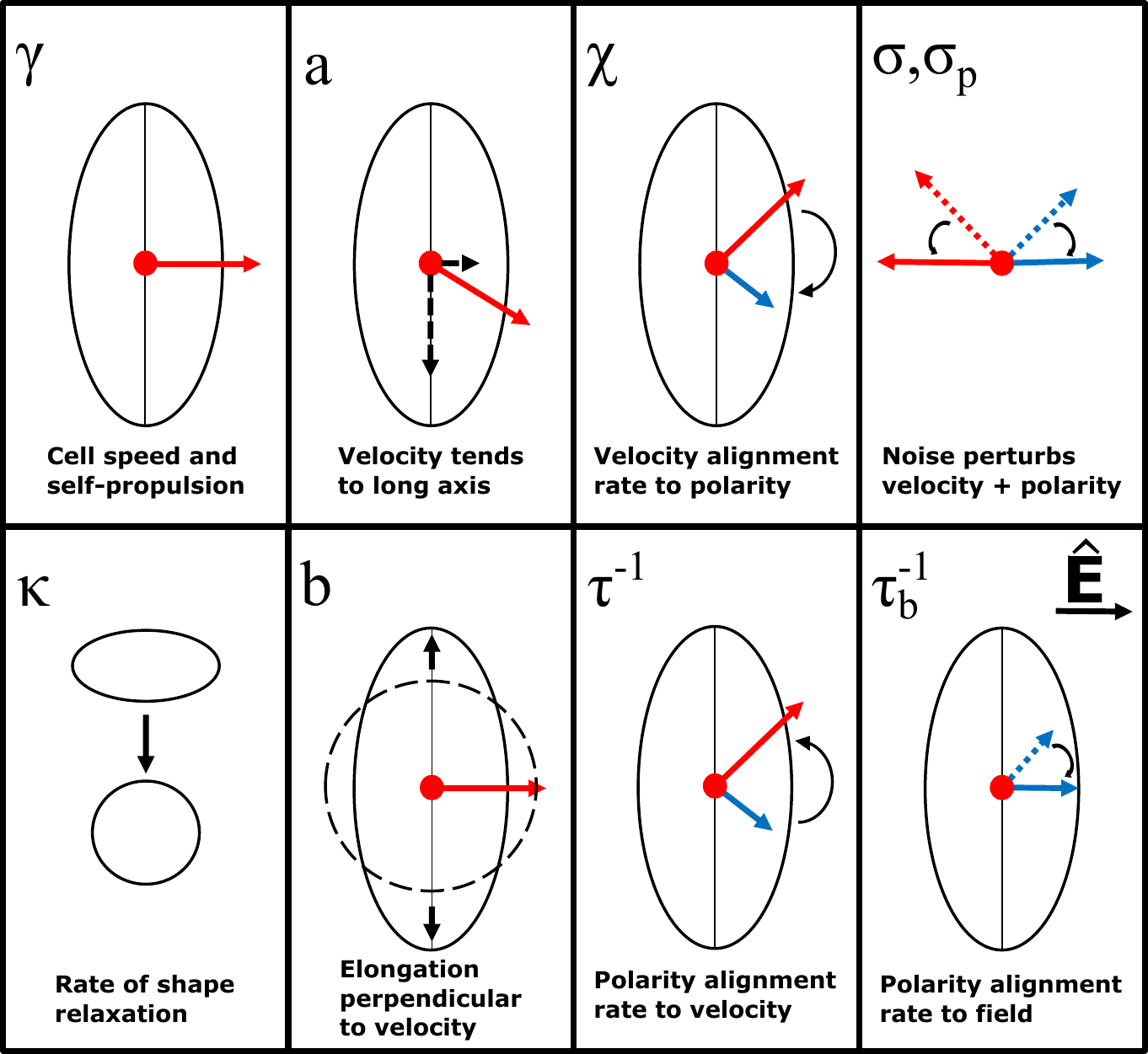}
	\caption{Schematic of the different terms in the equations of motion for velocity (Eq. \ref{eq:velocity}), shape (Eq. \ref{eq:shape}), and polarity (Eq. \ref{eq:polarity}).}
	\label{fig:Panel}
\end{figure}

Our model's dynamics for $S_{\alpha\beta}$, $\mathbf{v}$, and $\hat{\mathbf{p}}$ are given in Eqs. \eqref{eq:velocity}-\eqref{eq:polarity} and shown schematically in Fig. \ref{fig:Panel}. The cell's velocity obeys:
\begin{equation}
	\label{eq:velocity}
	\frac{\mathrm{d}\mathbf{v}}{\mathrm{d}t}=\mu(\gamma-\vert\mathbf{v}\vert^2)\mathbf{v}-a\dvec{\mathbf{S}}\cdot\mathbf{v}-\dfrac{\chi}{\vert \mathbf{v}\vert}\mathbf{v}\times(\mathbf{v}\times\mathbf{\hat{p}})+\sigma\xi(t)\mathbf{\hat{v}}_\bot.
\end{equation}
The first two terms on the right hand side are from \cite{ohta2009deformable}. The first term describes the cell's self-propulsion, which will (in the absence of other effects) drive the cell's speed to $\sqrt{\gamma}.$ The second term on the right hand side shows how the velocity responds to cell shape: if $a < 0$ (which is true everywhere in this paper), velocities along the cell's major axis will increase, and velocities along the cell's minor axis will decrease, leading the velocity to align to $\pm \hat{\mathbf{n}}$ (see Appendix \ref{app:shape-velocity}). The third term's triple cross product (a la \cite{czirok1996formation}) shows a rotation of velocity to come into alignment with the cell's polarity -- i.e. that the cell tries to move along its direction of internal biochemical polarity. This term is essential for galvanotaxis, as we assume the electric field alters the polarity $\hat{\mathbf{p}}$.  
The last term on the right is a rotational noise: $\xi(t)$ is a Gaussian Langevin noise with $\langle\xi(t)\rangle=0$ and $\langle\xi(t_1)\xi(t_2)\rangle=\delta(t_1-t_2)$.  $\mathbf{\hat{v}}_\bot$ is a unit vector perpendicular to $\mathbf{v}$. 

Cell shape $S_{\alpha\beta}$ obeys, following \cite{ohta2009deformable} exactly:
\begin{equation}
	\label{eq:shape}
	\frac{\mathrm{d}S_{\alpha\beta}}{\mathrm{d}t}=-\kappa S_{\alpha\beta}-
	b\left(v_\alpha v_\beta-\frac{1}{2}v^2 \delta_{\alpha\beta}\right).
\end{equation}
The first term in this equation informs us that elongated shapes tend to relax back to a circular shape with a rate $\kappa$ -- similar to the role of cell tension in \cite{camley2017crawling}. The second term on the right of Eq. \ref{eq:shape} models cell shape responding to cell velocity, and was proposed by \cite{ohta2009deformable} as the simplest term obeying the symmetries of the problem. If $b < 0$, cells expand their shape perpendicular to the direction of the velocity and contract parallel to the velocity, and vice versa if $b>0$. As keratocytes have their long axis perpendicular to the direction of travel, we will have $b < 0$ throughout this paper.

Lastly, the time evolution of the angle of the polarity, $\mathbf{\hat{p}}=(\cos{\phi_{\mathrm{p}}},\sin{\phi_{\mathrm{p}}})$, is given by:
\begin{equation}
	\label{eq:polarity}
	\frac{\mathrm{d}\phi_{\mathrm{p}}}{\mathrm{d}t}=-\dfrac{1}{\tau}\arcsin{[(\mathbf{\hat{v}}\times\mathbf{\hat{p}})_z]}-\dfrac{1}{\tau_{\mathrm{b}}}\arcsin{[(\mathbf{\hat{E}}\times\mathbf{\hat{p}})_z]}+\sigma_{\mathrm{p}}\xi_{\mathrm{p}}(t).
\end{equation}
The first term here aligns the cell's polarity with its velocity over a timescale $\tau$ i.e. a cell will want to polarize along its direction of travel \cite{szabo2006phase,camley2014velocity,szabo2010collective,basan2013alignment,camley2017physical}. The second term similarly shows that the polarity tends to align to the electric field's direction $\hat{\mathbf{E}}$ over a timescale $\tau_{\mathrm{b}}$. This is the only role of the electric field within our model.  The term $\xi_{\mathrm{p}}$ is a Gaussian Langevin noise with zero mean and $\langle \xi_{\mathrm{p}}(t_1) \xi_{\mathrm{p}}(t_2) \rangle = \delta(t_1-t_2)$. Within our model, we only include the electric field's direction -- we do not explicitly include its magnitude, which could alter the timescale over which the polarity aligns with the bias, $\tau_{\mathrm{b}}$. 

\section{\label{sec:results}Results}

\begin{figure*}[ht]
	\includegraphics[width=0.8\textwidth]{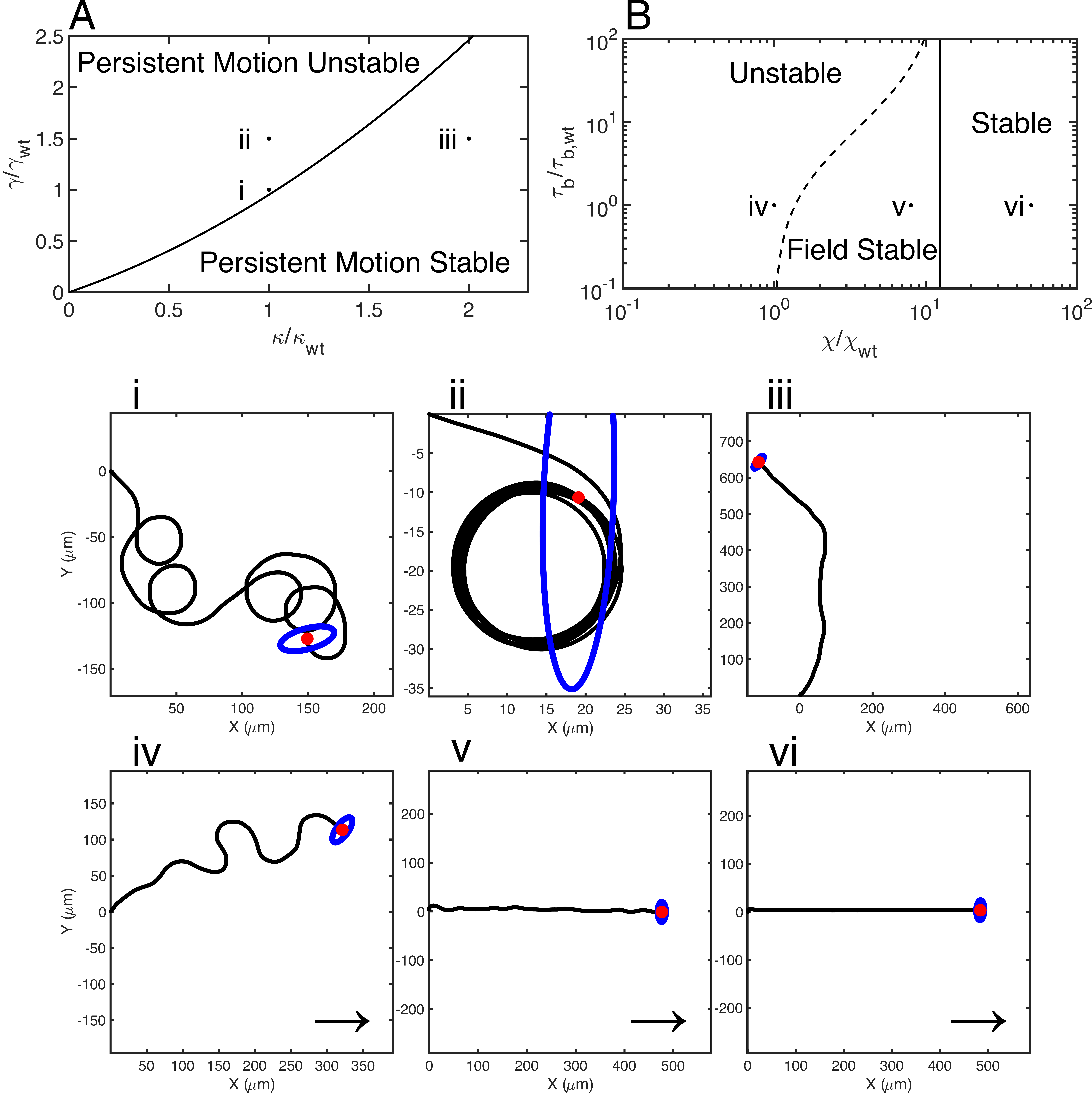}
	\centering
	\caption[FIGURE 2]{\textbf{(A)}: Linear stability phase diagram in $\gamma-\kappa$ plane in the absence of electric field. Example trajectories for points $i-iii$ are shown below. \textbf{(B)}: Linear stability phase diagram in $\tau_{\mathrm{b}}-\chi$ plane. Transition lines in presence of field (solid line) and absence (dotted line) are shown, and examples $iv, v, vi$ are shown with field.  \textbf{(i)}: shows a 2.5-hour trajectory at the ``wild-type" (WT) values, ($\gamma=\gamma_{\mathrm{wt}},~\kappa=\kappa_{\mathrm{wt}}$). \textbf{(ii)}: 1-hour trajectory when cell speed is increased slightly, resulting in persistent circular motion, ($\gamma=1.5\gamma_{\mathrm{wt}},~\kappa=\kappa_{\mathrm{wt}}$). \textbf{(iii)}: 1.5-hour trajectory with greatly increased shape relaxation rate $\kappa$ shows persistent random walk, ($\gamma=1.5\gamma_{\mathrm{wt}},~\kappa=2\kappa_{\mathrm{wt}}$). \textbf{(iv)}: 1.5-hour WT trajectory following a field in $+\hat{\mathbf{x}}$-direction. In this region of the plot, persistent motion is unstable, but the cell oscillates about the field direction. \textbf{(v)}: 1.5-hour trajectory following an field pointing in $+\hat{\mathbf{x}}$-direction. In this region of the plot, persistent motion is stabilized in the presence of an field and there are no oscillations, ($\tau_{\mathrm{b}}=\tau_{\mathrm{b,wt}},~\chi=8\chi_{\mathrm{wt}}$). \textbf{(vi)}: 1.5-hour trajectory following an field pointing in $+\hat{\mathbf{x}}$-direction, ($\tau_{\mathrm{b}}=\tau_{\mathrm{b,wt}},~\chi=50\chi_{\mathrm{wt}}$).}
	\label{fig:PhaseContour}
\end{figure*}

\subsection*{Phase diagram of model in absence of a field}
\label{subsec:lsa}
We show the phases of cell motion in the absence of a field in Fig. \ref{fig:PhaseContour}A.   With no field present and no noise, we observe two relevant steady states: cells may either travel in a straight line with their long axis perpendicular to their velocity, or move in a circular trajectory. Like the original work of \cite{ohta2009deformable}, we observe a transition between perfectly straight, linear motion at small cell speeds (low $\gamma$) and large shape relaxation rates (large $\kappa$) and circular motion at large $\gamma$ and small $\kappa$. We can identify this transition line by a linear stability analysis around the steady-state of a cell crawling in a straight line, neglecting noise (Appendix \ref{app:linearstability}). We find two transition lines $\gamma_1$ and $\gamma_2$, 
\begin{align} \label{eq:gamma1}
	\gamma_1&=\dfrac{\kappa^2}{ab}+\left(\dfrac{1}{2\mu}+\dfrac{\chi+\tau^{-1}}{ab}\right)\kappa+\dfrac{\chi+\tau^{-1}}{2\mu}\\[10pt]
	\label{eq:gamma2}
	\gamma_2&=\dfrac{\chi\tau+1}{ab}\kappa^2+\dfrac{\chi\tau+1}{2\mu}\kappa.
\end{align}
Linear motion is stable to perturbations if $\gamma < \textrm{min}(\gamma_1,\gamma_2)$. Once $\gamma>\gamma_1~\textrm{or}~\gamma_2$, then linear motion is unstable. We mark the region of the phase diagram Fig. \ref{fig:PhaseContour}A where linear motion is stable as ``Persistent Motion Stable."

\begin{figure}[ht]
	\includegraphics[width=0.52\textwidth, height=5.5cm]{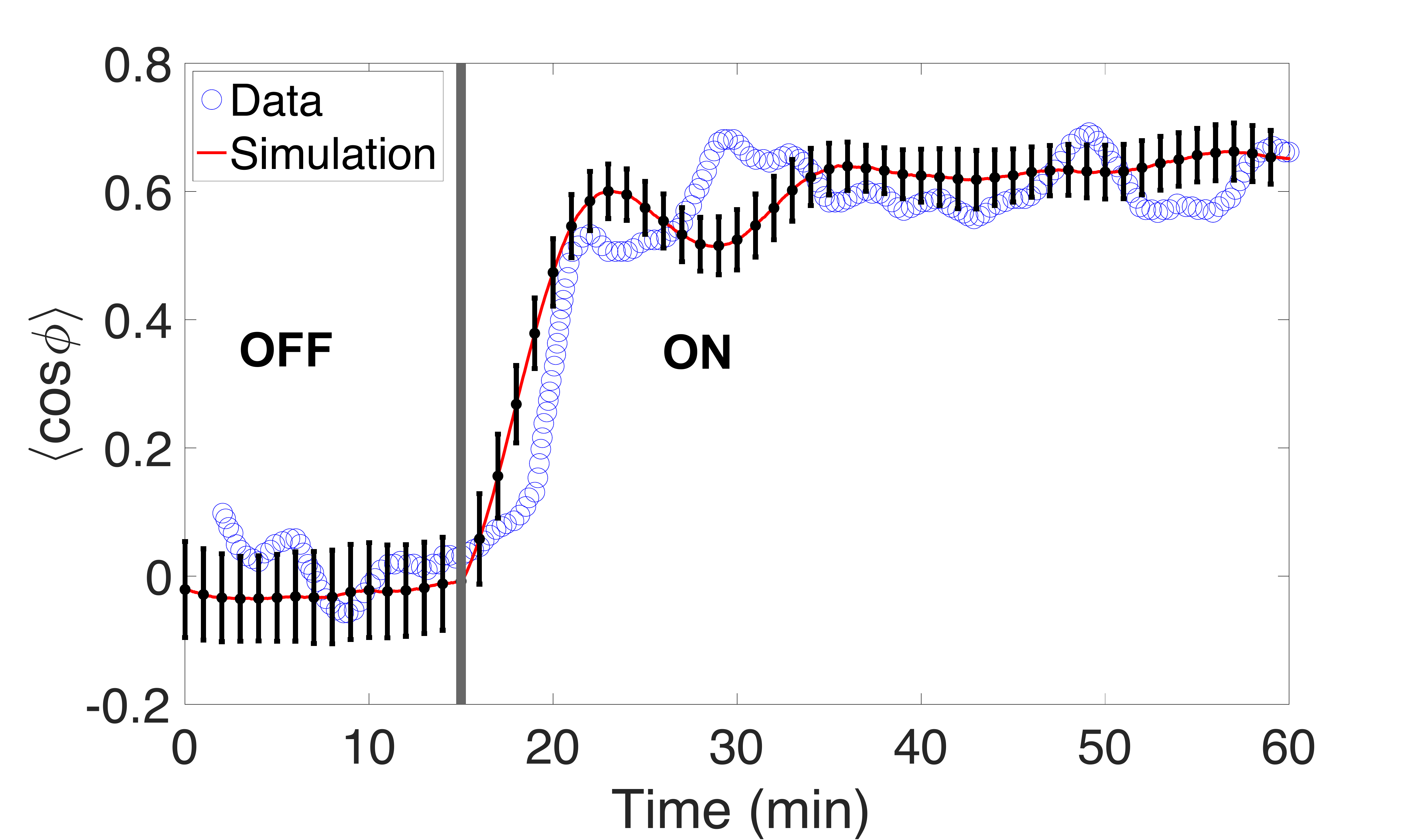}
	\centering
	\caption{Cell response to a field being turned on. Here, $\phi$ is the angle of the cell's velocity with respect to the field direction, and the field is turned on at 15 minutes. In blue is experimental data of the average response of 140 cells, extracted from \cite{allen2013electrophoresis}. In red is our simulation data which we fit to experiment to calibrate our parameters, averaging over 1000 cells. Error bars are $\pm 1$ standard deviation of the average, computed by bootstrapping (Section \ref{sec:methods}).}
	\label{fig:Data}
\end{figure}

The phase diagram is plotted with respect to the ``wild-type" parameters, the default fitted values for our model parameters. We found these parameters from fitting our model to data from \cite{allen2013electrophoresis} showing cells' response to a field being turned on (see Section \ref{sec:methods}: Methods and Fig. \ref{fig:Data})

In Fig. \ref{fig:PhaseContour}A, we highlight three points on the phase diagram and show example trajectories. The curve in Fig. \ref{fig:PhaseContour}A is simply $\gamma_2$, because $\gamma_2 < \gamma_1$ in the region of parameter space presented. Unlike the phase diagram, which was constructed without noise, the example trajectories have noise. In the presence of noise, the orientation of the cell changes even in the linear motion phase, and the cell undergoes a persistent random walk due to its orientational fluctuations. Thus, we define linear motion as characterized by perfectly straight cell migration without noise. Meanwhile, persistent motion is the analog of linear notion in the presence of noise, characterized by a persistent random walk with no loops, turns, or oscillations. We see in Fig. \ref{fig:PhaseContour}A that our wild-type values, represented by point $i$, are slightly above the transition line, where linear/persistent motion is unstable; meaning that the cell would exhibit circular motion/noisy trajectory with loops, turns, and oscillations. At the wild-type parameters, we see cell trajectories that resemble \enquote{knots on a string} \cite{allen2020cell}, where the cell goes through episodes of circular movement (Fig. \ref{fig:PhaseContour}i, Movie S1). The wild-type values being right above the transition line is due to an assumption in our fitting process: because both circular and persistent motion are observed in normal keratocytes, reasonable parameters must be close to the transition (Section \ref{sec:methods}). 

As we move away from the transition line and increase cell speed by increasing $\gamma$ (Fig. \ref{fig:PhaseContour}ii, Movie S2), we see that the cell travels perpetually in a circle, slightly perturbed by noise. If we move across the transition by increasing cell shape relaxation rate $\kappa$ (Fig. \ref{fig:PhaseContour}iii, Movie S3), we see that persistent migratory motion is stabilized again. While the boundary derived from Eqs. \eqref{eq:gamma1} and \eqref{eq:gamma2} is determined without added noise, it nonetheless captures a key transition in the model between persistent and circular migration.

\subsection*{Field stabilizes cell response}
Fig. \ref{fig:PhaseContour}A shows cell behaviors in the absence of a field. Fig. \ref{fig:PhaseContour}B shows how these phases can change in the presence of a field. In our simulations, we noticed that unstable linear motion without a field can stabilize with a field (Movie S4), depending on where we are in parameter space. We extended our linear stability analysis to the case with a field present with no noise (Appendix \ref{app:linearstability}). The linear stability curves are shown in the lines in Fig. \ref{fig:PhaseContour}B, where the solid line demarcates the transition between unstable linear motion and stable linear motion in the {\it absence} of an electric field, and the dotted line demarcates the transition in the {\it presence} of an electric field. However, even if the straight trajectory is linearly unstable, the cell may still be able to follow the field. Fig. \ref{fig:PhaseContour}iv shows an example of a trajectory from the phase diagram using our wild-type parameters. (This trajectory, like all our plotted example trajectories, includes the stochastic noise terms).  In Fig. \ref{fig:PhaseContour}iv, the field is on, but persistent motion is unstable. This is apparent by the presence of loops, turns, and oscillations. It is crucial to note that cells in the regime of globally unstable linear motion, such as Fig. \ref{fig:PhaseContour}iv, clearly can still follow a field. The linear stability analysis shows whether the linear trajectory following the field is stable to small perturbations -- and at the wild-type parameters, it is not. However, once we move across the solid transition line by increasing the rate $\chi$ that $\mathbf{v}$ aligns with $\mathbf{{\hat{p}}}$, linear motion is unstable only in the absence of a field, meaning that a cell would migrate persistently without oscillations when a field is present. We then see the oscillations around the field direction disappear in Fig. \ref{fig:PhaseContour}v. Once we cross the dotted transition line by increasing the rate of $\mathbf{v}-\mathbf{\hat{p}}$ alignment further, Fig. \ref{fig:PhaseContour}vi shows us a cell following the field even more persistently.

\subsection*{Simulations recapitulate experimental response to field}
To validate our model parameters and determine their wild-type values, we fit our simulations to data from Allen {\it et al}. (2013) \cite{allen2013electrophoresis}.  This data, shown in blue dots in Fig. \ref{fig:Data}, shows the experimental data of the average response of 140 cells responding to a field being switched on after 15 minutes of no exposure. We simulated 1000 cells freely traveling, responding to a field pointing in the $+\mathbf{\hat{x}}$-direction that is switched on after 15 minutes. In Fig. \ref{fig:Data}, we plot the average response of those 1000 cells as directedness vs. time in minutes. The directedness is defined by $\langle\cos{\phi}\rangle$, where $\phi$ is the angle of the cell velocity relative to the field. For the first 15 minutes when no field is present, we observe cells on average to go in all directions, so $\langle\cos\phi\rangle\approx 0$. Once the field turns on, the cells in the experiment respond to the field and follow it to varying levels of precision. In the figure, we see the directedness leveling off at $\langle\cos{\phi}\rangle\approx0.6$. This reflects both orientational noise and cell trajectories that often oscillate around the field direction (Fig. \ref{fig:PhaseContour}iv). Details of the fit are discussed in Section \ref{sec:methods}.  

Our model captures the experimentally observed response to a field being switched on reasonably well. What parameters influence this response and to what degree?

\begin{figure*}[ht]
	\includegraphics[width=0.8\textwidth]{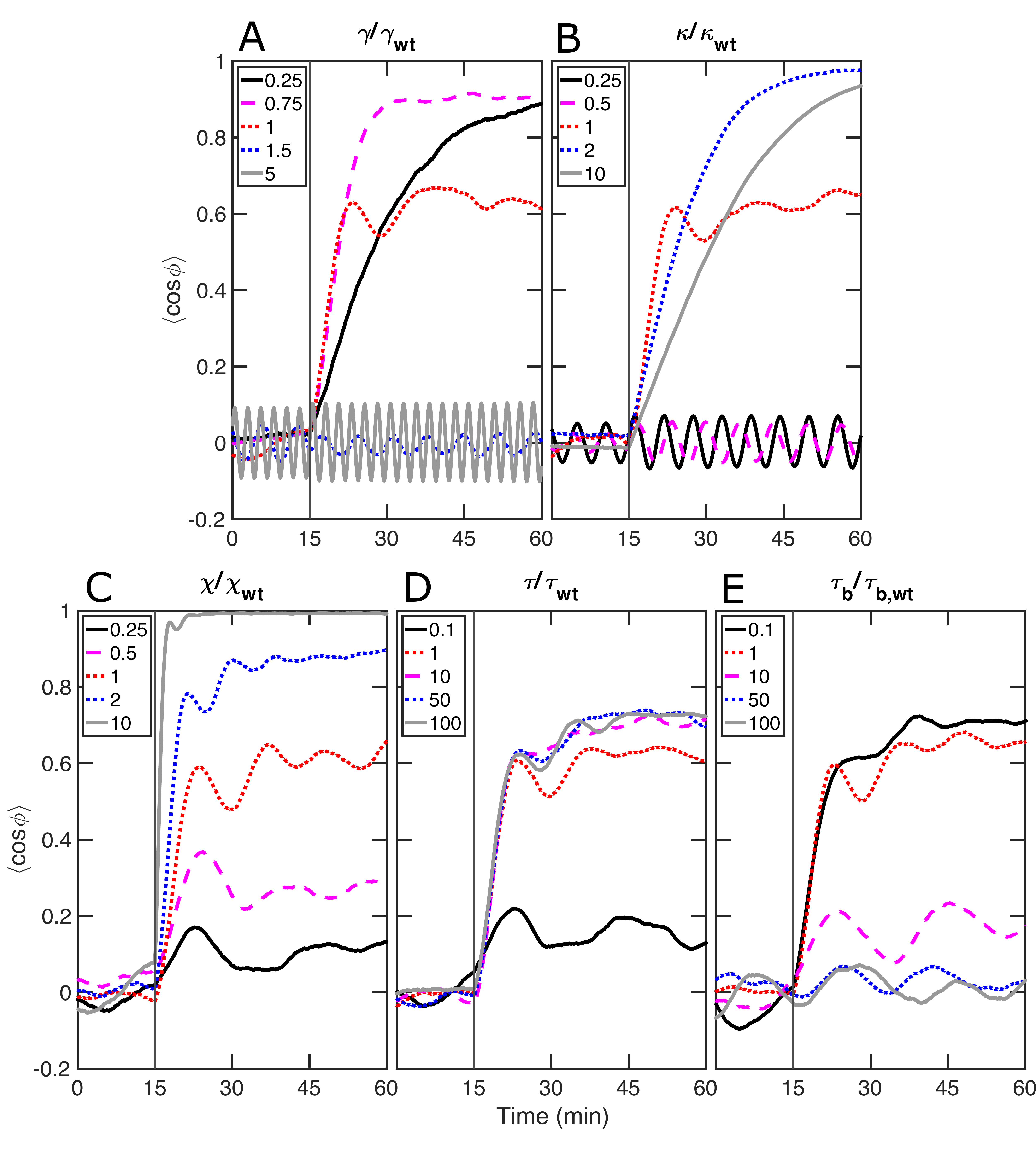}
	\centering
	\caption[FIGURE 4]{Response to a field being turned on at 15 minutes, as in Fig. \ref{fig:Data}, with parameters varied from fit wild-type parameters. Parameters varied are \textbf{(A)}: speed parameter $\gamma$. \textbf{(B)}: shape relaxation rate $\kappa$. \textbf{(C)}: velocity to polarity alignment rate $\chi$. \textbf{(D)}: polarity to velocity alignment time $\tau$. \textbf{(E)}: field alignment time $\tau_{\mathrm{b}}$. Each curve is the average response of 500 cells. Error bars are not shown in order to preserve clarity, but curve-to-curve variability is limited (see dotted red wild-type curve from panel to panel).}
	\label{fig:SummaryCurves}
\end{figure*}

\subsection*{Cell speed, shape, and ability to align with field  control cell response to field activation}
In Fig. \ref{fig:SummaryCurves} we use the cell directedness $\langle \cos \phi \rangle$ in response to a field being turned on in order to understand the influence of model parameters. We see that cells with lower speeds, $\gamma < \gamma_{\mathrm{wt}}$, have a more reliable response -- their value of $\langle \cos \phi \rangle$ saturates to a value of $\sim 0.9$, compared to $\sim 0.6$ for the wild-type cells (Fig. \ref{fig:SummaryCurves}A). However, this may come at a cost: for a slow enough cell, $\gamma = 0.25 \gamma_{\mathrm{wt}}$, the response to a field turned on at 15 minutes takes over 45 minutes to reach its steady-state level. This behavior is closely linked to where the cell's parameters are in the phase diagram (Fig. \ref{fig:PhaseContour}). For speeds much above the wild-type speed, e.g. $\gamma = 1.5 \gamma_{\mathrm{wt}}$, cells undergo a mostly circular trajectory and there is no net directedness. Based on our linear stability results, we would expect that increasing the rate of shape relaxation $\kappa$ would make linear, persistent motion more stable, allowing for a more directed response. This is what we find (Fig. \ref{fig:SummaryCurves}B): increasing $\kappa$ leads to a more directed response (saturation value of $\langle \cos \phi \rangle$ nearing 1). Similarly to our results for cell speed, increasing shape relaxation rate $\kappa$ can both increase the accuracy and slow the response time (see $\kappa = 10 \kappa_{\mathrm{wt}}$ curve). 

In our model, the electric field reorients cell polarity $\hat{\mathbf{p}}$ -- but for the cell's actual motility to reflect this change in polarity, the coupling between polarity and velocity is essential. The rate of this coupling is $\chi$. We see that if we decrease $\chi$ below the wild-type value, the response to signal becomes systematically less directed (Fig. \ref{fig:SummaryCurves}C). However, increasing $\chi$ can simultaneously increase accuracy and the response time of cells -- increasing $\chi$ to $10 \chi_{\mathrm{wt}}$ results in a cell that reaches near-perfect alignment with the field in under five minutes. This suggests that, with our fit parameters, bringing the velocity in alignment with the polarity can be the slowest element of cell reorientation. This is consistent with the small value of the time scale $\tau_{\mathrm{b}}$: polarity aligns with the field in much less than a minute. 

We have also modeled an alignment of cell polarity back to cell velocity (``self-alignment" \cite{szabo2006phase,camley2014velocity,camley2014polarity,camley2017physical}), which occurs with timescale $\tau$. Self-alignment appears to not be essential to cell response. Increasing $\tau$ (reducing the influence of self-alignment) even by orders of magnitude does not have a large effect on directionality (Fig. \ref{fig:SummaryCurves}D). By contrast, if $\tau$ is decreased to the order of $\tau_{\mathrm{b}}$, e.g. at $\tau=0.1\tau_{\mathrm{wt}}$, the cell's polarity must compromise between the field direction and the current velocity, and the cell responds more slowly and has a low directedness saturation level. 

Our model has the cell polarity align to the field over a timescale $\tau_{\mathrm{b}}$, subject to stochastic noise. Increasing $\tau_{\mathrm{b}}$ should then reduce the influence of the electric field. By increasing $\tau_{\mathrm{b}}$ sufficiently ($\tau_{\mathrm{b}} = 50 \tau_{\mathrm{b,wt}}$) we can nullify cell response to the field entirely (Fig. \ref{fig:SummaryCurves}E). Because $\tau_{\mathrm{b}}$ controls cell repolarization to the field, we would naively expect decreasing $\tau_{\mathrm{b}}$ below the wild-type to make the cell response to field turn-on faster. However, instead we see little effect of setting $\tau_{\mathrm{b}} = 0.1 \tau_{\mathrm{b,wt}}$, suggesting that the reorientation of polarity to the field is not rate-limiting -- sensible as $\tau_{\mathrm{b,wt}} = 0.024$ minutes, far less than the timescale for cell reorientation. The relative unimportance of $\tau_{\mathrm{b}}$ is consistent with the importance of $\chi$: solely increasing $\chi$ can make cells respond in $<5$ minutes.

\begin{figure*}[ht]
	\includegraphics[width=0.8\textwidth]{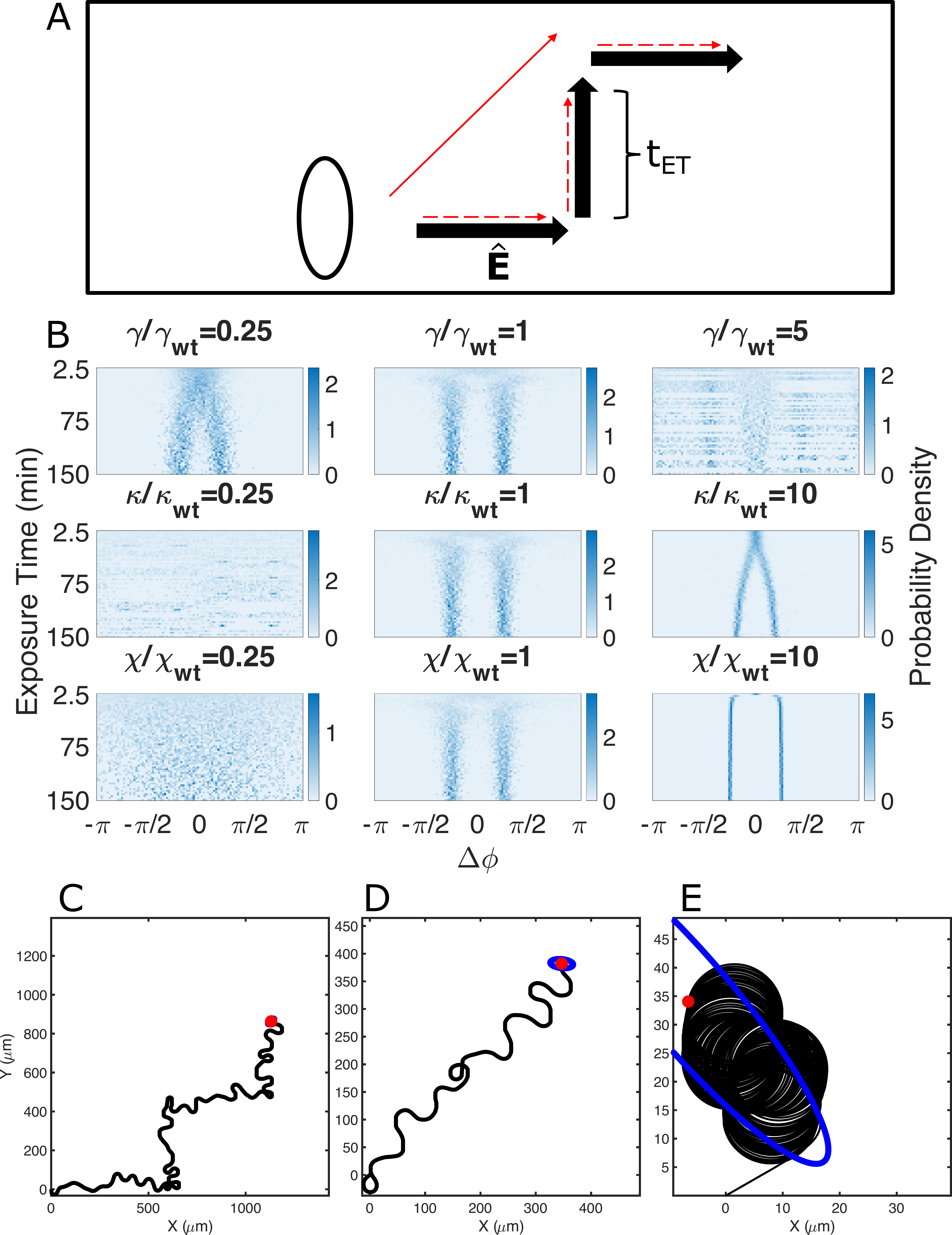}
	\centering
	\caption[FIGURE 5]{\textbf{(A)}: Schematic of simulation. We vary the electric field direction from $+\hat{\mathbf{x}}$ to $+\hat{\mathbf{y}}$ every $t_{\mathrm{ET}}$ and track the cell trajectory. Will the cell follow the field closely (dotted red arrows) or average the field directions (solid red arrow)? \textbf{(B)}: Probability density plots. These are histogram heatmaps indicating the probability of a cell averaging and pointing in the composite direction of the two field directions (+$\mathbf{\hat{x}}$ and +$\mathbf{\hat{y}}$) versus pointing in the field directions directly. Simulations are done for 100 simulation hours. \textbf{(C)}: 10-hour wild-type trajectory with slow switching time (2.5 hours), \textbf{(D)}: 3-hour wild-type trajectory with fast switching time (5 minutes), \textbf{(E)}: 4-hour trajectory with high $\gamma$ value where cell rotates faster than our averaging sample time.}
	\label{fig:StaircasePlots}
\end{figure*}

\subsection*{Cell response to switching fields includes averaging, tracking, and more complex behaviors}
Our results in Fig. \ref{fig:Data} and \ref{fig:SummaryCurves} show how a cell responds to a field that is switched on and kept in a fixed direction. Fig. \ref{fig:StaircasePlots} illustrates the results we see when we observe how a cell responds to a field being continually switched between the $+\hat{\mathbf{x}}$- and $+\hat{\mathbf{y}}$-directions. This is motivated by the experiments of Zajdel {\it et al}. \cite{zajdel2020scheepdog}, who probed collective cell migratory response to a rapidly switching electric field (Fig. \ref{fig:StaircasePlots}A). They discovered that on short time scales (20 seconds) the group of cells time average the varying field and follow the composite direction. For example, if  a field switches from $E_0 \hat{\mathbf{x}}$ and $E_0 \hat{\mathbf{y}}$, cells travel in a 45 degree angle. However, on longer times, cells follow the field direction reliably. 

When single cells respond to time-varying fields, do they track it closely (Fig. \ref{fig:StaircasePlots}C, Movie S5) or average the field (Fig. \ref{fig:StaircasePlots}D, Movie S6)? The experimental results of \cite{zajdel2020scheepdog} and our results of Fig. \ref{fig:SummaryCurves} suggest that when a field is switched on, there is a delay in response, which affects how well a cell would track or average a field. We know some of the parameters that control this. To what degree do these parameters control responses to a changing field?? On what time scale is a switching field considered \enquote{rapid}?  
 
In Fig. \ref{fig:StaircasePlots}, we show how cells respond to a switching electric field. During each simulation, we vary the direction of the field from $+\hat{\mathbf{x}}$ to $+\hat{\mathbf{y}}$ every  $t_{\mathrm{ET}}$ minutes -- the \enquote{exposure time} (Fig. \ref{fig:StaircasePlots}A). $t_{\mathrm{ET}}$ is held constant throughout a single simulation. We show results for how the cell behavior varies with exposure time, varying exposure time over 60 values from $t_{\mathrm{ET}}=2.5$ minutes to 2.5 hours. We show in Fig. \ref{fig:StaircasePlots}B the cell's response to a switching field as a function of exposure time by showing the probability density of the cell velocity's angle $\phi_{\mathrm{cell}}$. We plot this angle relative to the reference angle of 45$^\mathrm{o}$, $\phi_{\mathrm{r}}=\pi/4$, which the cell would take if the cell averaged the signal perfectly. (We compute the angles of the velocity averaged over the exposure time; see Appendix \ref{app:staircase}). We anticipated that a cell would take the average of a dynamically changing field at short exposure times, when the cell could not adequately respond in the timescale of the changing field. In this case, the cell should travel along $\phi_{\mathrm{r}}=\pi/4$. However, when the exposure time is much longer than the cellular response time (typically on the order of 10 minutes), we would expect that the cell tracks the changing field direction, producing a staircase-like trajectory. Fig. \ref{fig:StaircasePlots}B shows heat maps to illustrate this result. Here, $\Delta\phi=\phi_{\mathrm{cell}}-\phi_{r}$. A cell traveling along $\phi_{\mathrm{r}}=\pi/4$ would produce a heat plot with density centered at $\Delta\phi=0$. However, as the exposure time increases, we would expect to see two distinct bands at $\Delta\phi=\pm\pi/4$, a sign that $\phi_{\mathrm{cell}}=\{0,\pi/2\}$, indicating that the cell is traveling in a staircase pattern (Fig. \ref{fig:StaircasePlots}C). This characteristic shape is seen most clearly in Fig. \ref{fig:StaircasePlots}B when $\kappa=10\kappa_{\mathrm{wt}}$. This is consistent with the response to a field turning on at $\kappa=10\kappa_{\mathrm{wt}}$ seen in Fig. \ref{fig:SummaryCurves}B, where we see a very high level of cell directedness, but a slow response. Hence, we expect cells would not be able to follow a field with a short switching time, but would track reliably at long exposure times, forming solid bands at $\pm\pi/4$. However, this branching response is not as evident in the other parameters shown in Fig. \ref{fig:StaircasePlots}B. The heat map for $\chi=10\chi_{\mathrm{wt}}$ is similar, but shows two clear bands at nearly all exposure times except the shortest ones -- there is no averaging unless the field is switched less than every 5 minutes. Again, this is consistent with Fig. \ref{fig:SummaryCurves}C, where high values of $\chi$ have a response time of $<5$ minutes and a highly directed response. $\gamma=0.25\gamma_{\mathrm{wt}}$ has a heatmap with a forking pattern, but the noise blurs the bands since the cell does not follow and respond to the field as strongly, consistent with Fig. \ref{fig:SummaryCurves}A.
 
The middle panels of Fig. \ref{fig:StaircasePlots}B all show heatmaps at the wild-type parameter values, however, like $\gamma=\gamma_{\mathrm{wt}}0.25$, the heatmaps look blurred at short exposure times, though there are clear peaks at $\pm \pi/4$ at longer exposure times. The behavior at short exposure time shows a broad distribution of angles $\Delta \phi$, but with a peak at zero, indicating that the cell is most commonly going in the expected direction $\phi_{\mathrm{r}}$. Fig. \ref{fig:StaircasePlots}C and D show representative examples of trajectories at the wild-type parameters when the field is switched at 2.5 hours and 5 minutes, respectively (see also Movies S5 and S6). 
 
Even with a field on, we observe some parameter values where cells move in circles, which leads to strange results in the corresponding heatmaps in Fig. \ref{fig:StaircasePlots}B. For $\gamma=5\gamma_{\mathrm{wt}}~\text{and}~\kappa=0.25\kappa_{\mathrm{wt}}$, we do not see a branching pattern nor do we see peaks at $\Delta \phi = \pm \pi/4$, as the cell is rotating -- which can be seen in  Fig. \ref{fig:SummaryCurves}A and B as oscillations in the directionality. This rapid rotation does follow a sort of staircase pattern (Fig. \ref{fig:StaircasePlots}E, Movie S7) -- but does not result in a net movement in the direction of the fields.  In this regime, the rotating cell body moves perpendicular to the field direction. Depending on the direction of the cell's rotation (clockwise or counter-clockwise), the cell will then move in different directions in the field. In this example where the cell rotates counterclockwise, the cell's center of mass drifts in the $+\hat{\mathbf{y}}$-direction when the field points in the $\hat{\mathbf{x}}$-direction, and the cell drifts in the $-\hat{\mathbf{x}}$-direction when the field points in $+\hat{\mathbf{y}}$. When the cell rotates clockwise, the drifts will be in the opposite direction. This rotational motion explains the patchy structure of the heat maps of $\gamma = 5 \gamma_{\mathrm{wt}}$ and $\kappa = 0.25 \kappa_{\mathrm{wt}}$: they are constructed from one long simulation, where the cell spontaneously breaks symmetry from its initial conditions, ending up rotating clockwise or counter-clockwise, setting its direction of net drift. Therefore, we see either a large density at $\Delta \phi > 0$ or $\Delta \phi < 0$, which changes from one exposure time to another. We see from this that the influence of our stochastic noise is attenuated at these extreme parameter values, as the cell's initial symmetry breaking persists over the length of the trajectory. 

For the parameter set $\chi = 0.25 \chi_{\mathrm{wt}}$, cells are much less effective at following the field (Fig. \ref{fig:SummaryCurves}C), and we do not see a strong directed response in the heat maps. This is also in the regime where cells are experiencing episodes of rotation, albeit not always in a perfectly circular pattern (Movie S8). However, the heat map is not as patchy as the heat maps for $\gamma = 5 \gamma_{\mathrm{wt}}$ and $\kappa = 0.25 \kappa_{\mathrm{wt}}$. Instead of seeing larger densities at $\Delta \phi > 0$ or $\Delta \phi < 0$, we see a more uniform distribution with slight densities near $\Delta \phi = \pi/4$ or $\Delta \phi = -\pi/4$ at longer exposure times. This is due to the slower rotation and less directed motion in this regime. Hence, this regime has a greater sensitivity to noise. We are plotting the direction averaged over the exposure time. For $\gamma = 5 \gamma_{\mathrm{wt}}$, $\kappa = 0.25 \kappa_{\mathrm{wt}}$, cells often are able to perform several circular motions per averaging period -- so the average angle picks up on the net drift. Slower rotations, as in $\chi = 0.25 \chi_{\mathrm{wt}}$, lead to less directed velocities. Therefore, we don't see  analogous patchy bands in the $\chi = 0.25 \chi_{\mathrm{wt}}$ case.

\section{\label{sec:discussion}Discussion}

Our model recapitulates many separate qualitative migration behaviors observed in single keratocyte cells in or out of a field, including straight linear motion, persistent turning, turning while following a field (Fig. \ref{fig:PhaseContour}iv), and more complex ``knots on a string'' dynamics (Fig. \ref{fig:PhaseContour}i). These behaviors occur at slightly different parameter points in our model, but they can be created from relatively small changes in parameters -- so we expect keratocytes, which have a broad distribution of different properties including speed and aspect ratio, would naturally exhibit all of the phases shown in Fig. \ref{fig:PhaseContour}. This results from our initial assumption that wild-type model parameters had to be near the transition calculated from the linear stability analysis. 

Within our results, we have only studied cells in the presence or absence of a field, and have not systematically varied the strength of the field, which is not explicitly invoked in our model. There is some evidence that stronger electric fields increases directedness and decreases response times of keratocytes \cite{cooper1986motility,sun2013keratocyte,allen2013electrophoresis}. However, these results may be difficult to interpret, as Allen $et~al$. \cite{allen2013electrophoresis} observed speed increases in keratocytes correlated with increasing field -- but also noted that the increase in voltage also led to an increase in temperature, which separately increases cell speed \cite{ream2003influences}. The most natural interpretation of our model to study field strength systematically is to assume that the rate at which the polarity aligns with the field, $\tau_{\mathrm{b}}^{-1}$, could increase as the field increases. For our fit parameters, we predict relatively small changes from making the time $\tau_{\mathrm{b}}$ smaller (Fig. \ref{fig:SummaryCurves}e) -- suggesting that increasing field strength beyond 1 V/cm (the field used in the experimental data shown in Fig. \ref{fig:Data}) would not directly increase taxis, but that dropping the field could lead to impaired directionality. It is also possible that electric fields could have effects on single cells beyond repolarization, such as an increase in cell speed (electrokinesis). Increasing field strength has been observed to increase cell speed in neural progenitor cells and 3T3 fibroblasts \cite{meng2011pi3k,finkelstein2004roles}, though
field strengths as low as 0.25 V/cm are enough to induce a statistically significant response in individual cells and cell clusters \cite{allen2013electrophoresis}. These possibilities have been systematically studied by Prescott $et~al$., in experiments on human corneal epithelial cells, who  found that the only consistent model was one where cells preferentially polarize
in the direction of the field \cite{prescott2021quantifying}, supporting our broad assumptions about the polarity.  Within our model, if cell speeds changed as a function of field strength, this could lead to dramatic changes in single-cell response (Fig. \ref{fig:SummaryCurves}A).

A core element of our model is coupling between cell motion and cell shape, which drives persistent circular motion and oscillation. This coupling could arise through the dynamics of Rho GTPases within the cell \cite{camley2017crawling,singh2021sensing}, and similar effects can be generated through actomyosin instabilities \cite{nickaeen2017free}. How can experimental interventions be interpreted in our minimal model without these elements? $\kappa$ is a suitable parameter for this interpretation. $\kappa$ is the rate of cell shape relaxation from an ellipse to a circle. Our model trajectories reveal that a cell with a slower rate of shape relaxation (transitioning from large $\kappa$ to $\kappa_{\mathrm{wt}})$ can hold an elongated shape for longer periods of time. Slower shape relaxation allows the cell to respond to cues and perturbations more quickly; however, it tends to sacrifice accuracy in following these cues and perturbations (Fig. \ref{fig:SummaryCurves}). We find that more elongated cells with small values of $\kappa$ turn more quickly (Fig. \ref{fig:SummaryCurves}). However, when $\kappa\ll\kappa_{\mathrm{wt}}$, cells begin to oscillate. Lee $et~al.$ observed that cells with myosin activity inhibited by blebbistatin (low contractility) were less likely to oscillate \cite{lee2020modeling}, developing a model in which unequal amounts of adhesion on the left and right sides of the cell driven by myosin contractility destabilize linear motion, causing the cell to oscillate. It would be natural to assume that our shape relaxation rate $\kappa$ would increase with contractility. This leads to a contradiction with \cite{lee2020modeling}, as we see oscillation at low $\kappa$, not high contractility. However, blebbistatin has other effects on cell motility, including changing cell speed \cite{barnhart2011adhesion}. In our model, reduction in cell speed stabilizes linear motion (Fig. \ref{fig:PhaseContour}). Thus, the blebbistatin treatments in \cite{lee2020modeling} may not be straightforward to predict.

Zajdel $et$ $al.$ \cite{zajdel2020scheepdog} studied how collections of MDCK kidney epithelial cell line and primary, neonatal mouse skin keratinocytes respond to a rapidly changing field. They saw that at exposure times of 10 seconds, a group of cells would average the field and follow the composite direction, but would follow a staircase pattern for longer exposure times of 1.5 hours. Our results are broadly in agreement with Zajdel $et$ $al.$ in the regime where persistent motion is stable. $\kappa=10\kappa_{\mathrm{wt}}$ is a very clear example. The \enquote{wishbone} pattern clearly shows that at small time scales, we see the cell averaging and taking the composite field, but at large time scales, we see a cell follow a staircase pattern. The branching point of the wishbone is determined by the cell response time, which is determined by the transition region in the directedness vs. time plots (Fig. \ref{fig:SummaryCurves}), the region where the graph has a sharp incline before saturating at steady value of $\langle \cos \theta \rangle$. Other heat maps show similar behavior, but the level of noise masks the effects. For example, in the wild-type, the density at short time scales is centered at zero, but has a wide spread. Understanding to what extent these single-cell response curves influence the ability of cell groups to turn around is a future direction.

One striking prediction of our model is that, in extreme parameter ranges where cells undergo rapid rotation, they will travel perpendicularly to the field, with a directionality that depends on the cell's direction of rotation (Fig. \ref{fig:StaircasePlots}E). In principle, this suggests it could be possible {\it in vitro} to use an electric field to separate cells that rotate clockwise versus counterclockwise. While we only see this behavior at extreme parameter values when stochasticity becomes less important, our results show that this is possible, highlighting that this behavior is allowed by the symmetry of the system, when the clockwise/counter-clockwise rotation breaks the symmetry between $\pm \hat{\mathbf{y}}$ with the field in the $\hat{\mathbf{x}}$-direction. This drift of single cells is similar to behavior observed in simulations of clusters of rotating cells following a chemical gradient \cite{camley2016collective}.


\section{\label{sec:methods}Methods}

\subsection*{Model calibration and fitting}

\begin{table*}[ht]
	\begin{tabular}{|l|l|l|p{7cm}|l|}
		\hline
		\textbf{Parameter} & \textbf{Value} & \textbf{Units}                         & \textbf{Description}                                                & \textbf{Source}   \\ \hline
		$v_0$              & 10             & $\mu\mathrm{m~min}^{-1}$               & characteristic keratocyte velocity                                  & \cite{jurado2005slipping,keren2008mechanism}        \\ \hline
		$\gamma$           & $v_0^2$        & $\mu\mathrm{m}^2~\mathrm{min}^{-2}$    & propulsion term                                                     & Defined as $v_0^2$        \\ \hline
		$\kappa$           & 1.3068         & min$^{-1}$                             & shape relaxation rate                                               & fit \\ \hline
		$\chi$             & 0.0720         & min$^{-1}$                             & rate of velocity aligning with polarity                             & fit \\ \hline
		$\tau$             & 0.0617         & min                                    & time scale of polarity aligning with velocity                       & fit \\ \hline
		$\tau_{\mathrm{b}}$           & 0.0238         & min                                    & time scale of polarity aligning with bias                           & fit \\ \hline
	$\omega_0$         & 0.9758         & min$^{-1}$                             & 	characteristic frequency of keratocyte rotation, defined as $\omega_0 = \mu v_0^2$                     & Fit constrained by \cite{allen2020cell}\\ \hline 
		$\sigma$           & 0.3372         & $\mu\mathrm{m~min}^{-1.5}$             & velocity rotational noise                                           & fit \\ \hline
		$\sigma_{\mathrm{p}}$         & 0.7058         & min$^{-1.5}$                           & polarity displacement noise                                         & fit \\ \hline
		$a$                & -0.1132        & $\mu\mathrm{m}^{-1}\mathrm{~min}^{-1}$ & controls rate of velocity responding to shape ($a<0$ means velocity along cell's long axis increases) & fit \\ \hline
		$b$                & -0.5423        & min $\mu\mathrm{m}^{-1}$               & controls rate of shape responding to velocity   ($b<0$ means cell shape expands perpendicular to velocity direction)      & fit \\ \hline
		$\Delta t$           & 0.001         & min                                    & time step for simulations                           &  \\ \hline
	\end{tabular}
\caption{Table of parameter values, description, and sources.}
    \label{tab:pvalues}
\end{table*}

Where possible, we use well-known cell parameters for shape, size, and speed. We choose a characteristic speed of $v_0 = 10 ~\mu\mathrm{m~min}^{-1}$ \cite{jurado2005slipping,keren2008mechanism}, which fixes $\gamma = v_0^2$. We note that the actual speed of the cell will depend on other parameters other than $\gamma$ (see Appendix \ref{app:linearstability}); we have only set its rough order of magnitude by choosing $v_0$.

To set the remainder of our parameters, we fit our model to the experiments results of Allen $et~al$. \cite{allen2013electrophoresis}, who studied cell directionality in response to an electric field being turned on (shown in Fig. \ref{fig:Data}). We extracted this data with WebPlotDigitizer \cite{Rohatgi2020}. The experiments show that if the cell's velocity has an angle $\phi$ to the $x$-axis, the average directionality $\langle \cos \phi \rangle$ increases to a steady state value over a time of $\sim 5-10$ minutes after the field is turned on (Fig. \ref{fig:Data}). 

The parameters that needed to be calculated were $a, b, \kappa, \omega_0,\chi,\tau,\tau_{\mathrm{b}},\sigma$, and $\sigma_{\mathrm{p}}$; we reparameterized $\mu=\omega_0/v_0^2$. 
To get our fit parameters, we picked from a broad range of plausible values (See Appendix \ref{app:range}) using Latin hypercube sampling, and then ran a simulation to match the experiment by having no electric field when $t<15$ min, then setting $\mathbf{{E}}=[1,0]$ when $t\geq15$ min. To minimize the influence of initial conditions, we simulate 2 hours of trajectory as equilibration time before the $t = 0$ point of the trajectories shown in Fig. \ref{fig:Data}. We ran 1000 repeated simulations to construct the average directionality $\langle \cos \phi \rangle$ over time. The same procedure of equilibration and averaging was conducted for the curves in Fig. \ref{fig:SummaryCurves}, but with 500 simulations. Parameter sets that led to numerical instabilities were discarded; we found these largely corresponded to cells oscillating unphysically fast. We then chose the parameters that created the best-fitting directionality curve, according to our criterion (see Eq. \ref{eq:fit_error} below).

We know that in order to observe both circular motion and persistent motion, as seen experimentally, we must have our parameters close to the transition between circular and straight motion (see Section \ref{sec:results}). To ensure we are near this transition point, and to reduce the number of fit parameters, we constrained the possible fit values for $b$ and $\kappa$ using two pieces of information: 1) the rough experimental aspect ratio of keratocytes, and 2) the idea that the cells must be near the transition between circular and persistent motion. The value of the shape deformation parameter $s$ emerges from the model; if the cell is at steady state in a straight line, it takes on the value 
\begin{equation*}
   s_{\mathrm{ss}} = -\frac{b}{\kappa} \frac{\mu \gamma}{\beta + \mu},~\text{where}~\beta = \frac{a b}{2 \kappa}.
\end{equation*}
Note that $a,b<0$. The values $\kappa$ and $b$ that would be found {\it if the parameters were at the transition line and at steady state} are
\begin{align*}
\kappa_1&=-a s-(\chi+\tau^{-1}),\quad \kappa_2=\dfrac{-as}{\chi\tau+1}\\[10pt]
	b_{\mathrm{T}}&=\dfrac{2s\mu\kappa}{2\gamma\mu-as},
\end{align*}
where $s=s_{\mathrm{ss}}$ is the value of $s$ in the steady-state with the cell crawling straight (Appendix \ref{app:linearstability}). $b_{\mathrm{T}}$ is derived by rearranging $v^2_{\mathrm{ss}}$ (Appendix \ref{app:linearstability}). There are two equations for $\kappa$. Choosing $\kappa = \kappa_1$ sets the system to the transition line in the phase diagram if $\gamma_1 < \gamma_2$, and choosing $\kappa = \kappa_2$ sets the system to the transition line if $\gamma_2 < \gamma_1$. For all the plausible range of parameters shown in Table \ref{tab:ranges}, $\gamma_2 < \gamma_1$ and so we always choose $\kappa$ to be near $\kappa_2$ (Table \ref{tab:ranges}). In setting $\kappa$ and $b$, we assume $s_{\mathrm{ss}}=12.5 \mu \mathrm{m}$, which sets the aspect ratio of the keratocyte. We know experimentally that the roughly elliptical keratocyte cells of interest that we were modeling had an radius of $R_0=14.5~\mu\mathrm{m}$ for an equivalent circle of the same area. We then choose the steady state value of $s=12.5~\mu\mathrm{m}$, which ensures that the aspect ratio of the cell, $(R_0+s/2)/(R_0-s/2)\approx2.5$. 

Our goal is to find the parameter set that minimized the error between simulations and experiment -- but we suspect from our simulations that many of the oscillations about the steady-state value in the experimental data of Fig. \ref{fig:Data} are from the finite sample size of 140 cells. To avoid overfitting, we have chosen not to directly minimize the error between simulation and experiment in $\langle \cos \phi \rangle$ (Fig. \ref{fig:Data}). Instead, we want our model to correctly capture the response time and steady-state directionality of the data. We fit both the simulated and experimental measurements of $\langle \cos \phi \rangle$ to a hyperbolic tangent function:
\begin{equation}
	\langle \cos \phi \rangle=A\tanh{\left(\dfrac{t-t_{\mathrm{s}}}{\tau_{\mathrm{r}}}\right)}+B. \label{eq:fitform}
\end{equation}
using MATLAB's nlinfit. As described above, the simulation curves of $\langle \cos \phi \rangle$ were constructed from averaging together 1000 simulations of the field being turned on. In this fit function, $t_{\mathrm{s}}$ is the time of the inflection point of $\tanh{(\cdot)}$ i.e. the time where the derivative is maximal. $\tau_{\mathrm{r}}$ is the scaling factor that determines how \enquote{wide} the function is at the transition. Physically, $\tau_{\mathrm{r}}$ sets the scale of the time the cell takes to respond to an electric field being turned on. The other key parameter in the fitting form Eq. \eqref{eq:fitform} is the long-time value of the directionality, $SS = \lim_{t\to\infty} \langle \cos \phi \rangle = A + B$. We found the values $\tau_\textrm{expt}$ and $SS_\textrm{expt}$ from fitting the experimental curve to Eq. \eqref{eq:fitform}. We then chose, among the range of parameters, the parameters that minimize the error $\epsilon$ in steady-state value $SS=A+B$ and response time $\tau_{\mathrm{r}}$ between the experiment and theory:
\begin{equation}
	\epsilon = \left(\dfrac{\tau_{\mathrm{r}}-\tau_{\mathrm{expt}}}{\tau_{\mathrm{expt}}}\right)^2+\left(\dfrac{SS-SS_{\mathrm{expt}}}{SS_{\mathrm{expt}}}\right)^2 \label{eq:fit_error}
\end{equation}

The resulting wild-type parameters are shown in Table \ref{tab:pvalues}.

The errors bars in Fig. \ref{fig:Data} were calculated with the bootstrap method. 100 of the 1000 simulated cells were sampled and the mean of their response was calculated. This process was repeated for 100 samples. The standard deviation of the mean responses of the 100 samples were calculated and used as the error bars. 

\subsection*{Numerical methods}

We simulate our model equations (Eq. \eqref{eq:velocity}-\eqref{eq:polarity}) using the Euler-Maruyama method. To handle the noise $\xi(t)$ and $\xi_{\mathrm{p}}(t)$ numerically, we defined a new Gaussian distributed random variable $\Gamma$:
\begin{equation*}
	\Gamma(\Delta t)=\int_{t}^{t+\Delta t}\xi(t)~\mathrm{d}t.
\end{equation*}
Using the delta-correlation of the noise, $\langle\Gamma\rangle=0$ and $\langle\Gamma^2\rangle=\Delta t$. That means numerically, $\Gamma=\mathcal{X}\sqrt{\Delta t}$, where $\mathcal{X}\sim\mathcal{N}(0,1)$. For an example of this approach, we integrate Eq. \eqref{eq:velocity} from time $t$ to $t+\Delta t$:
\begin{equation*}
	\mathbf{{v}}(t+\Delta t)-\mathbf{{v}}(t)= f(\mathbf{{v}}(t),\dvec{\mathbf{S}}(t),\mathbf{\hat{p}}(t))\Delta t+\sigma\Gamma\mathbf{\hat{v}}_\bot(t),
\end{equation*}
Other equations are integrated similarly.

\begin{acknowledgments}
We thank Emiliano Perez Ipi\~na for a careful reading of the paper. We acknowledge support from NSF PHY 1915491 and NSF MCB 2119948. IN was supported by the Program of Molecular Biophysics 5T32GM135131-02.
\end{acknowledgments}

\onecolumngrid


\appendix

\section{Interpreting shape-velocity coupling}
\label{app:shape-velocity}
In our equations of motion for the velocity, we have a term $\dvec{\mathbf{S}}\cdot\mathbf{v}$, which is a matrix product, i.e. its $\alpha$th component is $S_{\alpha\beta}v_\beta$. 

We can write
\begin{equation}
	\dvec{\mathbf{S}}=s
	\begin{pmatrix}
		\cos^2{\theta}-\dfrac{1}{2} &  \cos{\theta}\sin{\theta} \\[10pt]
		\cos{\theta}\sin{\theta} & \sin^2{\theta}-\dfrac{1}{2}
	\end{pmatrix}=\dfrac{s}{2}
	\begin{pmatrix}
		\cos{2\theta} &  \sin{2\theta} \\[10pt]
		\sin{2\theta} & -\cos{2\theta}
	\end{pmatrix}.
\end{equation}

The eigenvectors of $\dvec{\mathbf{S}}$ are just the unit vectors parallel to the axis, $\hat{\mathbf{n}} = (\cos \theta,\sin\theta)$ and perpendicular to it, $\hat{\mathbf{n}}_\perp = (-\sin\theta,\cos\theta)$, with eigenvalues $\pm s/2$, respectively. If we break up the velocity into the components perpendicular to and parallel to the long axis, $\mathbf{v} = v_\parallel \hat{\mathbf{n}} + v_\perp \hat{\mathbf{n}}_\perp$, then 
\begin{align*}
\dvec{\mathbf{S}}\cdot\mathbf{v} &= \dvec{\mathbf{S}}\cdot(v_\parallel \hat{\mathbf{n}} + v_\perp \hat{\mathbf{n}}_\perp) \\
& = \frac{s}{2} v_\parallel \hat{\mathbf{n}} - \frac{s}{2}v_\perp \hat{\mathbf{n}}_\perp
\end{align*}
Thus, revisiting Eq. \eqref{eq:tensor}, we see that $-a\dvec{\mathbf{S}}\cdot{\mathbf{v}}=-as(v_\parallel \hat{\mathbf{n}} - v_\perp \hat{\mathbf{n}}_\perp)/2$. When $a<0$, as is true everywhere in this paper, this term then makes the velocity component along the long axis $\hat{\mathbf{n}}$ increase while the component along the short axis decreases -- this term makes the velocity align more along the long axis of the cell.

\section{Parameter range variation}
\label{app:range}

We show in Table \ref{tab:ranges} the ranges of parameter values used to fit in our Latin hypercube process.

\begin{table*}[ht]
\begin{tabular}{|l|l|p{9cm}|}
\hline
\textbf{Parameter} & \textbf{Range}                    & \textbf{Comments}                                                                                                                                                                                    \\ \hline
$\kappa$           & $[0.9,~1.1]\kappa_{1,2}$          & \begin{tabular}[c]{@{}l@{}}Linearly spaced\end{tabular}                                                                                         \\ \hline
$\chi$             & {[}0.01, 0.1{]}              & Logarithmically spaced                                                                                                                                                                               \\ \hline
$\tau$             & {[}0.01, 0.1{]}              & Logarithmically spaced                                                                                                                                                                               \\ \hline
$\tau_{\mathrm{b}}$           & $[0.01,~10]\tau$                  & Logarithmically spaced                                                                                                                                                                               \\ \hline
$\omega_0$         & $[0.9,~1.1]\dfrac{\pi}{3}$rad        & This constraint is both from the rough timescale of $\omega_0$ $\sim$ 1 deg/s  \cite{allen2020cell} and earlier fits, which pointed at this being the plausible range. \\ \hline
$\sigma$           & $[0.01,~0.1]\sqrt{v_0^2\omega_0}$ & Linearly spaced                                                                                                                                                                                      \\ \hline
$\sigma_{\mathrm{p}}$         & $[0.01,~0.1]\sqrt{v_0^2\omega_0}$ & Linearly spaced                                                                                                                                                                                      \\ \hline
$|a|$                & {[}0.1, 1{]}                 & Logarithmically spaced; $a$ is required to be negative so that the velocity along the long axis of the cell increases, leading to the turning instability (Appendix \ref{app:shape-velocity}).                                                                                                                                                                               \\ \hline
$b$                & $[0.9,~1.1]b_{\mathrm{T}}$                   & Linearly spaced;    $b$ is constrained to be near the value given by the transition condition (see Methods, Section \ref{sec:methods}). {$b$ is also required to be negative so that $ab>0$, which is necessary for the instability to occur \cite{ohta2009deformable}.}                                                                                                                                                                                 \\ \hline
\end{tabular}
\caption{Table showing range over which parameters were varied in the fit.}
\label{tab:ranges}
\end{table*}

\section{Linear stability analysis}
\label{app:linearstability}
We start by finding the stationary solutions of the deterministic portions of Eqs. \eqref{eq:velocity}-\eqref{eq:polarity}, which are a first step to finding linear stability but are also interesting in their own right. First, let us define $\mathbf{v}=v(\cos{\phi},\sin{\phi})$, where $v_1=v\cos{\phi}$ and $v_2=v\sin{\phi}$ are the first and second components, respectively. Earlier, we defined $\mathbf{\hat{n}}=(\cos{\theta},\sin{\theta})$ and $\mathbf{\hat{p}}=(\cos{\phi_{\mathrm{p}}},\sin{\phi_{\mathrm{p}}})$, meaning that $n_1$, $n_2$, $p_1$, $p_2$ can be defined the same way $v_1$ and $v_2$ were defined. Knowing the components of $\mathbf{\hat{n}}$ allow us to list the components of $\dvec{\mathbf{S}}$ using Eq. \eqref{eq:tensor}. 
Second, it will serve us to rewrite the deterministic portions of  Eqs. \eqref{eq:velocity} and \eqref{eq:shape} component-wise:
\begin{align*}
	\frac{\mathrm{d}v_\alpha}{\mathrm{d}t}&=\mu(\gamma- v^2)v_\alpha-aS_{\alpha\beta}v_\beta-\dfrac{\chi}{ v}(v_\alpha v_\beta-\vert v\vert^2\delta_{\alpha\beta})p_\beta,\\[10pt]
	\frac{\mathrm{d}S_{\alpha\beta}}{\mathrm{d}t}&=-\kappa S_{\alpha\beta}	+b\left(v_\alpha v_\beta-\dfrac{1}{2}\vert v\vert^2\delta_{\alpha\beta}\right).
\end{align*}
where we have implicitly assumed Einstein summation. After making the appropriate substitutions for the components of $\mathbf{v}$, $\mathbf{\hat{p}}$, and $\mathbf{\hat{n}}$, one can derive equations for the velocity magnitude, $v$, the angle of the velocity, $\phi$, the shape change $s$, and the angle of the major axis $\theta$ i.e. Eqs. \eqref{eq:comp1}-\eqref{eq:comp4}. To derive Eq. \eqref{eq:comp1}, one must calculate $\mathbf{\hat{v}}\cdot\partial_t\mathbf{v}=(\cos{\phi})\partial_t{v}_1+(\sin{\phi})\partial_t{v}_2$. To derive \eqref{eq:comp2}, one must calculate $\mathbf{\hat{v}}_\perp\cdot\partial_t\mathbf{v}=(-\sin{\phi})\partial_t{v}_1+(\cos{\phi})\partial_t{v}_2$. To derive Eqs. \eqref{eq:comp3} and \eqref{eq:comp4}, one must calculate $(\cos{2\theta})\partial_t{S}_{11}+(\sin{2\theta})\partial_t{S}_{12}$ and $(-\sin{2\theta})\partial_t{S}_{11}+(\cos{2\theta})\partial_t{S}_{12}$, respectively. These calculations yield the following equations:

\begin{align}
	\label{eq:comp1}
	\dfrac{\mathrm{d}v}{\mathrm{d}t}&=\mu(\gamma-v^2)v-\dfrac{asv}{2}\cos{2(\theta-\phi)}\\[10pt]
	\label{eq:comp2}
	\dfrac{\mathrm{d}\phi}{\mathrm{d}t}&=-\dfrac{as}{2}\sin{2(\theta-\phi)}-\chi\sin{(\phi-\phi_{\mathrm{p}})}\\[10pt]
	\label{eq:comp3}
	\dfrac{\mathrm{d}s}{\mathrm{d}t}&=-\kappa s+v^2b\cos{2(\theta-\phi)}\\[10pt]
	\label{eq:comp4}
	\dfrac{\mathrm{d}\theta}{\mathrm{d}t}&=-\dfrac{v^2b}{2s}\sin{2(\theta-\phi)}.
\end{align}
Following \cite{ohta2009deformable}, assuming that we have a steady state with the cell moving in the $+\hat{\mathbf{x}}$-direction, i.e. $\phi=0$, and that $ab>0$, the steady state solutions for the velocity magnitude, $v$, and the shape deformation, $s$, can be calculated by setting Eqs. \eqref{eq:comp1}-\eqref{eq:comp4} equal to 0 and solving. When $b<0$, the steady state occurs when $\theta = \pi/2$ and
\begin{equation*}
	s_{\mathrm{ss}}=-\dfrac{b}{\kappa}v_{\mathrm{ss}}^2,\quad v_{\mathrm{ss}}^2=\mu\dfrac{\gamma}{\beta+\mu},~\text{where}~\beta=\dfrac{ab}{2\kappa}.
\end{equation*}
(When $b>0$, the steady state occurs when $\theta = 0$ and $s_{\mathrm{ss}} = +\dfrac{b}{\kappa}v_{\mathrm{ss}}^2$.) 
$b<0$ is used everywhere in this paper for our model. These results so far are identical to the results of Ohta et al \cite{ohta2009deformable} except that we have introduced the scale $\mu$ to help make the units of the problem more explicit. At the steady state, $\phi = \phi_{\mathrm{p}}$, which removes the polarity's influence in Eq. \eqref{eq:comp2}; thus the polarity equation does not affect the steady-state values of $s$ or $\theta$.

Next, for linear stability analysis we can first consider the deterministic portion of Eq. \eqref{eq:polarity} without the influence of an electric field. In that case, Eq. \eqref{eq:polarity} simplifies to
\begin{equation}
    \label{eq:comp5}
	\frac{\mathrm{d}\phi_{\mathrm{p}}}{\mathrm{d}t}=-\dfrac{1}{\tau}\arcsin{(\mathbf{\hat{v}}\times\mathbf{\hat{p}})}=\dfrac{\phi-\phi_{\mathrm{p}}}{\tau}.
\end{equation}
where we have assumed that $|\phi-\phi_{\mathrm{p}}|$ is small enough that $\arcsin(\sin(\phi_{\mathrm{p}}-\phi)) = \phi_{\mathrm{p}}-\phi$, which will be true in our linear stability analysis below. Then, we can define two new variables, $\psi=\theta-\phi$ and $\epsilon=\phi-\phi_{\mathrm{p}}$, constructing two equations for our analysis from Eqs. \eqref{eq:comp1}-\eqref{eq:comp5}:
\begin{align}
	\label{eq:eps}
	\dfrac{\mathrm{d}\epsilon}{\mathrm{d}t}&=-\dfrac{as}{2}\sin{2\psi}-\chi\sin{\epsilon}-\dfrac{\epsilon}{\tau}\equiv f(\epsilon,\psi)\\[10pt]
	\label{eq:psi}
	\dfrac{\mathrm{d}\psi}{\mathrm{d}t}&=-\dfrac{1}{2}\left(-as+\dfrac{bv^2}{s}\right)\sin{2\psi}+\chi\sin{\epsilon}\equiv g(\epsilon,\psi).
\end{align}

Next, we have to linearize $f$ and $g$ with respect to $\epsilon$ and $\psi$ to first order. We evaluate the derivatives of $f$ and $g$ at $\epsilon^*=0$ and $\psi^*=\pi/2$ (i.e. at a steady state where velocity and polarity are pointed in the same direction, but the long axis of the cell is perpendicular to this, appropriate since $b<0$ for our model), giving us:
\begin{equation}
\frac{d}{dt} \begin{pmatrix}\epsilon \\ \psi \end{pmatrix} \approx     \mathbf{A} \begin{pmatrix}\epsilon \\ \psi \end{pmatrix} 
\end{equation}
where the Jacobian matrix
\begin{equation*}
    \mathbf{A}=
    \begin{pmatrix}
    	\dfrac{\partial f}{\partial\epsilon} &  \dfrac{\partial f}{\partial\psi} \\[10pt]
    	\dfrac{\partial g}{\partial\epsilon} &  \dfrac{\partial g}{\partial\psi}
    \end{pmatrix}
    =
    \begin{pmatrix}
        -\chi-\tau^{-1} & as \\[10pt]
        \chi & -as+\dfrac{bv^2}{s}
    \end{pmatrix}.
\end{equation*}
Here, $s=s_{\mathrm{ss}}$ and $v=v_{\mathrm{ss}}$. For $b>0$, the derivatives are evaluated at $\epsilon^*=0$ and $\psi^*=0$, altering the second column of the Jacobian matrix by a factor of -1.

For the solutions of our system of linearized equations to be stable, the eigenvalues of $\mathbf{A}$ must both have negative real parts, which requires  Tr$(\mathbf{A})<0$ and det$(\mathbf{A})>0$. We can derive Eqs. \eqref{eq:gamma1} and \eqref{eq:gamma2} by solving for the $\gamma$ values that set Tr$(\mathbf{A}) = 0$ and det$(\mathbf{A}) = 0$.

The linear stability results of our model  in the absence of a field reduces to that of Ohta \& Ohkuma \cite{ohta2009deformable} in the limit where $\mathbf{{v}}$ is no longer aligned with $\mathbf{\hat{p}}$ ($\chi\rightarrow0$), but $\mathbf{\hat{p}}$ is perfectly aligned with $\mathbf{{v}}$ ($\tau^{-1}\rightarrow\infty$) . In this case $\gamma_1\to\infty$ while $\gamma_2\to\gamma_{\mathrm{c}}=\dfrac{\kappa^2}{ab}+\dfrac{\kappa}{2\mu}$. As we noted above, $\mu = 1$ is chosen implicitly in \cite{ohta2009deformable}. 

We also conducted linear stability analysis with the electric field present in the deterministic portion of Eq. \eqref{eq:polarity}, which takes the form

\begin{equation}
	\frac{\mathrm{d}\phi_{\mathrm{p}}}{\mathrm{d}t}=\dfrac{\phi-\phi_{\mathrm{p}}}{\tau}+\dfrac{\phi_E-\phi_{\mathrm{p}}}{\tau_{\mathrm{b}}}=\dfrac{\phi-\phi_{\mathrm{p}}}{\tau}-\dfrac{\phi_{\mathrm{p}}}{\tau_{\mathrm{b}}}.
\end{equation}
$\phi_E=0$ since $\mathbf{E}=[1,0]$. To do stability analysis, we need three equations for $\epsilon$, $\psi$, and $\phi_{\mathrm{p}}$, as now the angle between the polarity and the electric field is important. $f(\epsilon,\psi,\phi_{\mathrm{p}})$ gains an extra term, $+\phi_{\mathrm{p}}/\tau_{\mathrm{b}}$, while $g(\epsilon,\psi,\phi_{\mathrm{p}})$ remains unchanged and we define $\partial_t{\phi_{\mathrm{p}}}\equiv h(\epsilon,\psi,\phi_{\mathrm{p}})$.
The resulting Jacobian matrix is
\begin{equation}
	\mathbf{A}=
	\begin{pmatrix}
		\dfrac{\partial f}{\partial\epsilon} &  \dfrac{\partial f}{\partial\psi} & \dfrac{\partial f}{\partial\phi_{\mathrm{p}}} \\[10pt]
		\dfrac{\partial g}{\partial\epsilon} &  \dfrac{\partial g}{\partial\psi} & \dfrac{\partial g}{\partial\phi_{\mathrm{p}}} \\[10pt]
		\dfrac{\partial h}{\partial\epsilon} &  \dfrac{\partial h}{\partial\psi} & \dfrac{\partial h}{\partial\phi_{\mathrm{p}}}
	\end{pmatrix}
	=
	\begin{pmatrix}
		-\chi-\tau^{-1} & as & \tau_{\mathrm{b}}^{-1}\\[10pt]
		\chi & -as+\dfrac{bv^2}{s} & 0\\[10pt]
		\tau^{-1} & 0 & -\tau_{\mathrm{b}}^{-1}
	\end{pmatrix}.
\end{equation}
where the derivatives are evaluated at the appropriate steady-state values of $\epsilon$, $\psi$, and $\phi_p$.

For the solutions to be stable, all three eigenvalues of $\mathbf{A}$ must have a negative real part. The analytical form for these conditions are cumbersome; we show stability diagrams determined by numerically finding the eigenvalues of $\mathbf{A}$ in Fig. \ref{fig:PhaseContour}B. 

\section{Probability density plots}
\label{app:staircase}
In Fig. \ref{fig:StaircasePlots}, we showed the effect of a rapidly switching fields in terms of the probability density of angles of the cell velocity.  In these simulations, the field was switched between the $+\hat{\mathbf{x}}$- and $+\hat{\mathbf{y}}$-directions with \enquote{exposure times} $t_\textrm{ET}$ ranging from every 2.5 minutes to 2.5 hours. The simulation was run for a total of $t_\textrm{max}=100$ hours. To measure the cell response, we calculated the angle between the average cell velocity in a given time window and the reference angle $\phi_{\mathrm{r}}=\pi/4$, which is the average direction of the changing field.\\
Algorithm:\\
\tab 1. Simulate the cell and find $\mathbf{v}(t)$ for a time $t_\textrm{ET}$, keeping the field constant. \\
\tab 2. Compute the average velocity over this time, $\bar{v}$\\
\tab 3. Find the angle between $\bar{v}$ and $\phi_{\mathrm{r}}$ and store: $\Delta\phi=\phi_{\textrm{cell}}-\phi_{\mathrm{r}}$, where $\phi_{\textrm{cell}} = \arctan{(\bar{v}_2/\bar{v}_1)}$\\
\tab 4. Switch the field direction and then repeat steps 1-3 until total time $t_\textrm{max}$. \\
\tab 5. Plot probability density of angles.

\section{Supplementary movie captions}
\label{app:figsmovies}
In all movies, the magnitude $|v|$ shows the magnitude of the velocity in units of $\mathrm{\mu m/min}$. Parameters not stated explicitly are given by the default values in Table \ref{tab:pvalues}.\\ \\
\textbf{Movie S1:} Trajectory of wild-type cell without any field.\\
\textbf{Movie S2:} Trajectory of cell at 1.5$\gamma_{\mathrm{wt}}$ without any field.\\
\textbf{Movie S3:} Trajectory of cell at 1.5$\gamma_{\mathrm{wt}}$ and 2$\kappa_{\mathrm{wt}}$ without any field.\\
\textbf{Movie S4:} Trajectory of cell at $8\chi_{\mathrm{wt}}$ with no noise. Field turns on after first 120 minutes.\\
\textbf{Movie S5:} Trajectory of wild-type cell with slowly switching electric field ($t_{\mathrm{ET}}=2.5$ hours).\\
\textbf{Movie S6:} Trajectory of wild-type cell with rapidly switching electric field ($t_{\mathrm{ET}}=5$ minutes).\\
\textbf{Movie S7:} Trajectory of cell at 5$\gamma_{\mathrm{wt}}$ with switching electric field ($t_{\mathrm{ET}}=80$ minutes). Cells rotate while showing drift perpendicular to field direction.\\
\textbf{Movie S8:} Trajectory of cell at 0.25$\chi_{\mathrm{wt}}$ with switching electric field ($t_{\mathrm{ET}}=80$ minutes). At these parameters, the polarity follows the field reliably but the cell direction does not. \\

\end{document}